\documentclass[a4paper]{spicawStyle}
\usepackage{graphicx,natbib}

\begin{document}

\title{The study  of debris disks with SPICA}

\author{Amaya Moro-Mart\'{\i}n}

\institute{
Department of Astrophysics, CAB (CSIC/INTA), 28850 Torrej\'on de Ardoz, Madrid, Spain\\
Department of Astrophysical Sciences, Princeton University, Princeton, NJ 08544, USA}

\maketitle 

\begin{abstract}

Debris disks are evidence that stars harbor reservoirs of dust-producing plantesimals on spatial scales  similar the solar system. Debris disks present a wide range of sizes and structural features and there is growing evidence that, in some cases, they might be the result of the dynamical perturbations of a massive planet. Our solar system also harbors a debris disk and some of its properties resemble those of extra-solar debris disks. This contribution discusses how the study of debris disks with SPICA can shed light on the diversity of planetary systems, the link between debris disks and planets and the link between extra-solar planetary systems and our own.
\keywords{circumstellar matter -- Kuiper Belt -- planetary systems}
\end{abstract}

\small
\begin{center}
1. DEBRIS DISKS ARE EVIDENCE OF THE PRESENCE OF PLANETESIMALS
\end{center}
\normalsize

Debris disks are disks of dust (with low gas-to-dust ratio) that surround mature main sequence stars with ages from 10 Myr to 10 Gyr. The maturity of the star is a critical issue. According to gas surveys, by 10 Myr the primordial gas in the protoplanetary disk has already dissipated and lifetime considerations \cite{backman1993} indicate that the primordial dust should also be gone\footnote{Dust particles are affected by Poynting-Robertson (P-R) drag, which in the  inertial reference frame can be thought of as the result of the particle emitting more (photon) momentum into the forward  direction of motion due to the Doppler effect. The particle loses momentum and since the mass is conserved,  the particle is decelerated and spirals slowly toward the central star (until it sublimates) in a timescale given by  $t_{PR} = 710\Big(\frac{s}{\mu m}\Big)\Big(\frac{\rho}{g/cm^3}\Big)\Big(\frac{R}{AU}\Big)^2\Big(\frac{L_{\odot}}{L_*}\Big)\frac{1}{1+albedo}$ years. Dust particles are also subject to {\it mutual collisions} that erode the grain away in a timescale given by $t_{col} = 1.26\cdot10^4\Big(\frac{R}{AU}\Big)^{3/2}\Big(\frac{M_{\odot}}{M_*}\Big)^{1/2}\Big(\frac{10^{-5}}{L_{dust}/L_*}\Big)$ years; within that time, the particles become smaller than the blowout size at which point they are blown away from the system by {\it radiation pressure}. Radiation pressure blowout happens in approximately an orbital period with $t_{blow} = \frac{1}{2}\Big(\frac{(R/AU)^3}{M_*/M_{\odot}}\Big)^{1/2}$ years.}.   This means that the dust in the debris disks originates from reservoirs of planetesimals -- similar to the asteroids, Kuiper Belt objects (KBOs) and comets in our solar system -- that release dust from mutual collisions, collisons with interstellar grains or sublimation. 
The fractional luminosity of the solar system debris disk is $L_{dust}/L_* \sim 10^{-7}-10^{-6}$ for dust located in the Kuiper Belt (KB) \cite{stern1996} and $L_{dust}/L_* \sim 10^{-8}-10^{-7}$ for dust produced in the asteroid belt (AB) \cite{dermott2002}, with effective dust temperatures of approximately 60 K and 200 K, respectively. 

\small
\begin{center}
2. FREQUENCY AND CHARACTERIZATION OF EXTRA-SOLAR PLANETESIMAL BELTS
\end{center}
\normalsize

\begin{figure}
  \includegraphics[angle = 0, height=0.215\textheight]{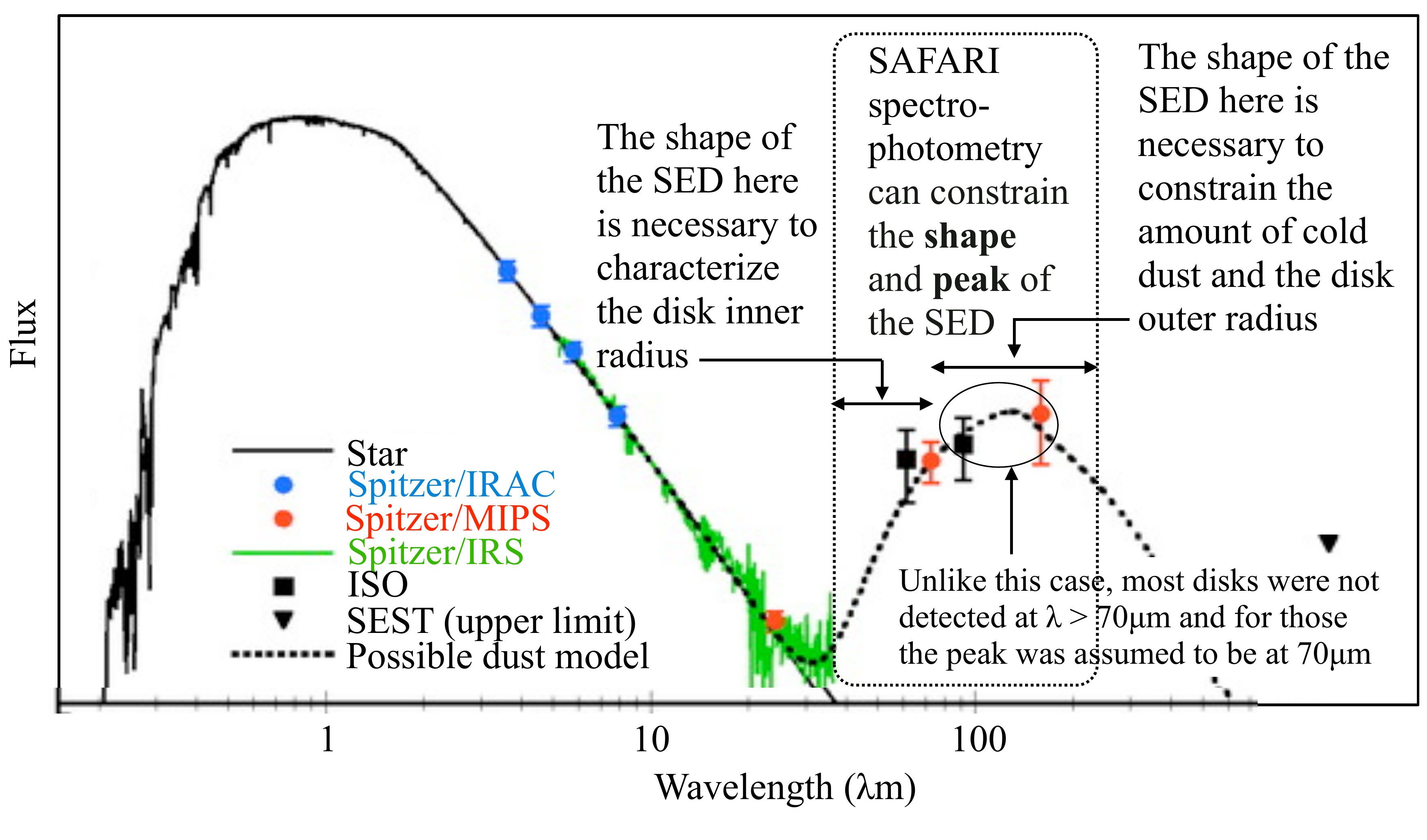}
  \caption{Spectral energy distribution of HD 105  \cite{meyer2004}. The lack of emission at  mid-infrared wavelengths is due to the depletion of dust at distances $<$ 35 AU from the central star. Most of the debris disks known to date have inner evacuated regions. The wavelengths covered by SAFARI (enclosed by the dotted lines) are critical to constrain the disk inner and outer radii.}
\label{sed}
\end{figure}

The great majority of the $\sim$ 300 debris disks known to date are spatially unresolved and have been identified from the presence of an infrared excess in the spectral energy distribution (SED)  with respect to what is expected from the stellar photosphere, thought to be due to the dust thermal emission.  For these spatially unresolved disks, we can estimate the dust location from the study of their SED:  different wavelengths trace different dust temperatures and therefore different dust locations; the warmer dust emitting at shorter wavelengths is located closer to the star than the colder dust emitting at longer wavelengths; if the warmer dust were to be absent (imagine a disk with a central cavity), the SED would have a depletion in the mid-IR region (see Figure \ref{sed}).

\begin{figure}
  \includegraphics[angle = 0, height=0.225\textheight]{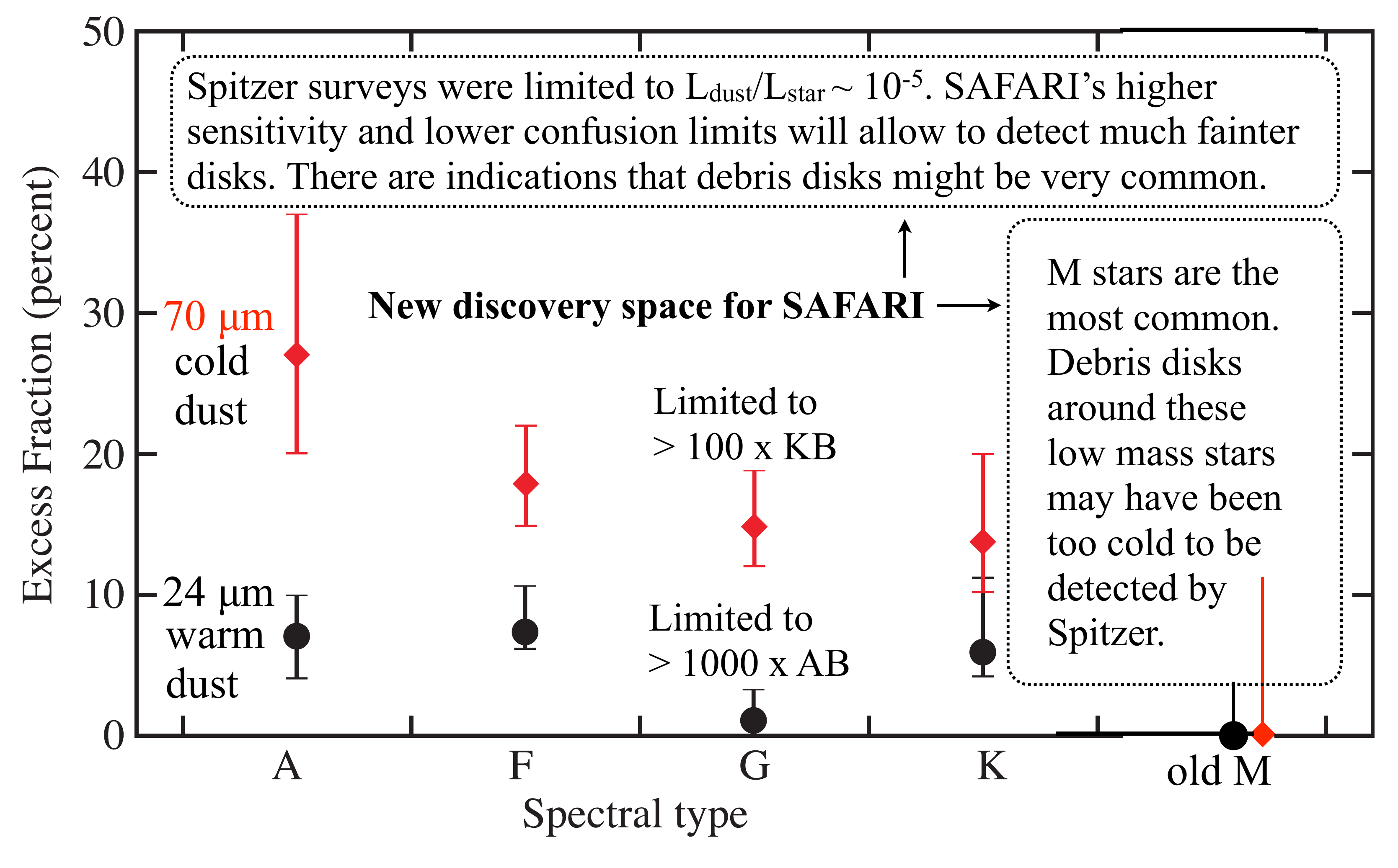}
  \caption{The percentage of stars with excess dust emission, i.e. with indirect evidence of the presence of dust-producing planetesimal belts, as a function of stellar type. These results are based on {\it Spitzer} observations. Figure adapted  from \cite{trilling2008} and  \cite{gautier2007}. The mean ages within each type are 0.7 Gyr, 3.8 Gyr, 5.7 Gyr and 6.0 Gyr for A, F, G and K-stars, respectively. The vertical bars correspond to binomial errors that include 68\% of the probability (1-$\sigma$ for Gaussian errors). The filled {\bf black} circles are for stars showing excess emission at 24 $\mu$m (tracing warmer dust) and empty {\bf red} diamonds are for stars showing excess emission at 70 $\mu$m (tracing colder KB-like dust). For solar G-type stars, {\it Spitzer} 70 $\mu$m debris disk detections were generally limited to $\sim$100$\times$ the luminosity of the dust in the Kuiper belt, while at 24 $\mu$m they were limited to $\sim$1000$\times$ the luminosity of the dust in the asteroid belt. The areas enclosed by the dotted lines (for fainter disks and disks around stars colder than K2) indicate SAFARI's discovery space.}
\label{incidence}
\end{figure}

Figure \ref{incidence} shows the debris disks incidence rates\footnote{For the 225 Sun-like (FG) stars older than 600 Myr in Figure \ref{incidence}, the frequency of the debris disks are 4.2$^{+2}_{-1.1}$\% at 24 $\mu$m and 16.4$^{+2.8}_{-2..9}$\% at 70 $\mu$m.  For comparison, the {\it Spitzer FEPS} survey of 328 FGK stars found that the frequency of 24 $\mu$m excess is 14.7\% for stars younger than 300 Myr and  2\% for older stars, while at 70 $\mu$m, the excess rates are 6--10\% \cite{hillenbrand2008},  \cite{carpenter2009}.} derived from a combined sample of 350 AFGKM stars \cite{trilling2008}. It is clear from the figure that cold dust emitting at 70 $\mu$m is more common than warm dust emitting at 24 $\mu$m. For solar type stars and assuming blackbody grains, this would correspond to dust located at 3 AU and 28 AU, respectively. Detailed analysis of the excess spectra (12--35 $\mu$m) of 44 FGK stars  shows that central cavities are common  with typical disk inner radius  of $\sim$ 40 AU and $\sim$ 10 AU for systems with and without 70 $\mu$m excess, respectively, confirming that most of the debris disks observed are KB-like \cite{carpenter2009}.

\begin{figure}
  \includegraphics[angle = 0, height=0.20\textheight]{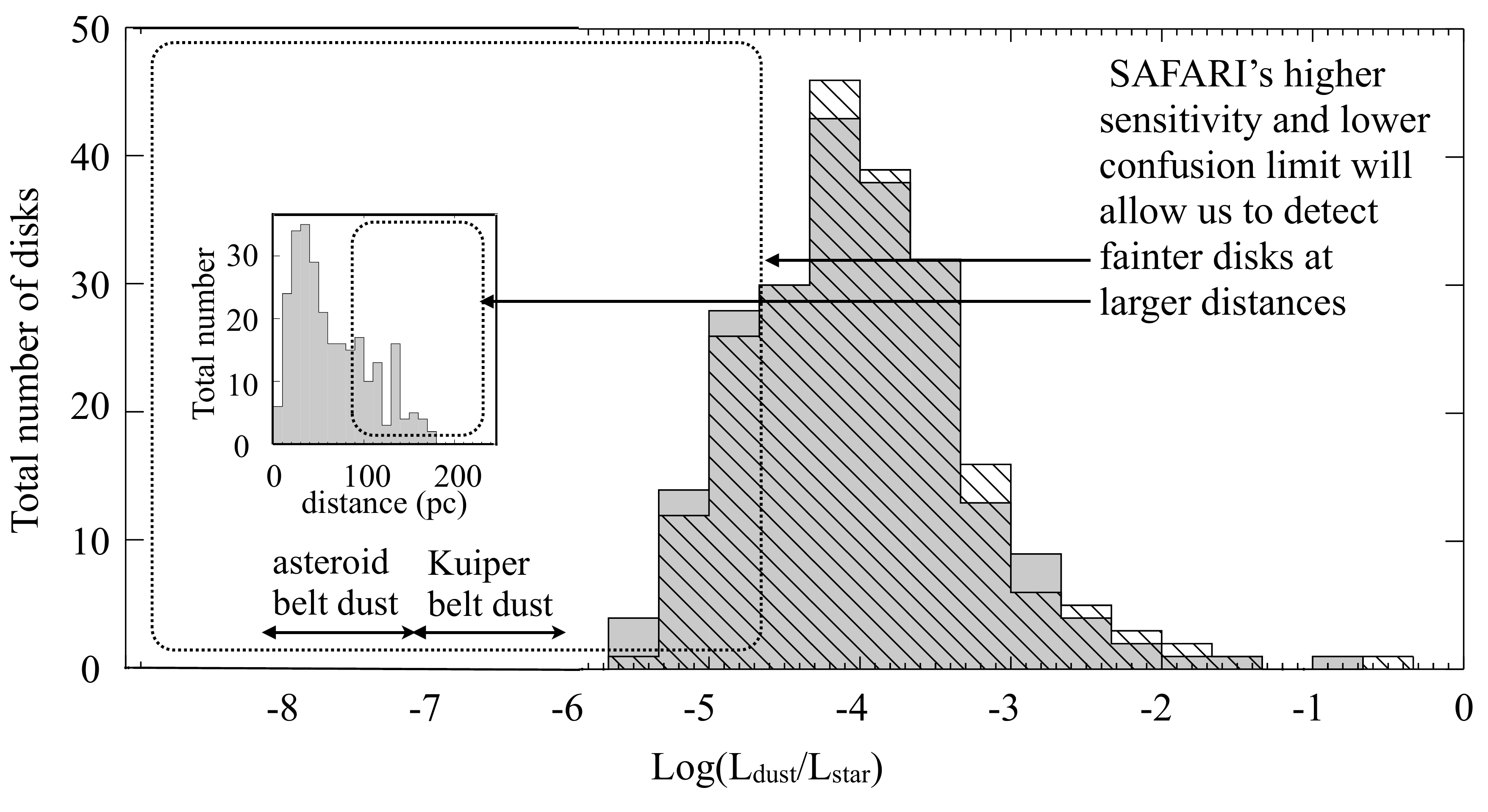}
  \caption{{\bf Main panel:} Distribution of fractional luminosities, L$_{dust}$/L$_*$, for the debris disks known to date (the y-axis is the total number of disks - not a percentage). Overlapping histograms indicate upper and lower limits. The characteristic values for the Kuiper and Asteroid belts are shown at the bottom left.  {\bf Insert:} Distribution of distances. The areas enclosed by the dotted lines indicate SAFARI's discovery space.}
\label{histo}
\end{figure}

The incidence rate for cold KB-like debris disks around solar-type stars  is similar to the $\sim$17--19\% of solar-type stars that harbor giant planets inside 20 AU \cite{cumming2008}.  However, the sensitivity of the {\it Spitzer} observations is limited to fractional luminosities of L$_{dust}$/L$_*$$>$10$^{-5}$ (see histogram in Figure \ref{histo}), i.e. $>$100 times the expected luminosity from the KB dust in our Solar system. Assuming a gaussian distribution of debris disk luminosities and 
extrapolating from {\it Spitzer} observations (showing that the frequency of dust detection 
increases steeply with decreasing fractional luminosity), it is found that the luminosity of the 
solar system dust is consistent with being 10 $\times$ brighter or fainter than an average 
solar-type star, i.e. debris disks at the solar system level could be common, but would have been too faint to be 
detected by {\it Spitzer} \cite{bryden2006}. Taking into account the limited sensitivity of the debris disks observations, it seems that 
systems harboring dust-producing KBOs are more common than those with giant planets. This is in 
agreement with the core accretion models  of planet formation where the planetesimals are the building blocks 
of planets and the conditions required to form planetesimals are less restricted than those to form gas giants. This also 
agrees with the observation that  the presence of giant planets is strongly correlated with high stellar metallicities \cite{fischer2005}, whereas debris disks are not \cite{greaves2006}, indicating that planetesimals can be formed under a broader range of conditions (also indicated by the two orders of magnitude difference in the luminosity of the stars that harbor debris disks), and with the observation that debris disks and planets are not correlated \cite{ama2007}. 

\small
\begin{center}
{\it Contribution of SPICA}
\end{center}
\normalsize

The scarcity of debris disks around stars later than K2 and around distant stars is likely an observational bias because these disks would have been too cold and too faint to be detected  by {\it Spitzer}. However, models and observations indicate that debris disks are probably very common and this large discovery space is yet to be explored. It is important to study the population of small bodies because it can give us a more complete picture of the diversity of extra-solar planetary systems, shedding light on their formation and dynamical histories. 

\begin{figure}
  \includegraphics[angle = 0, height=0.26\textheight]{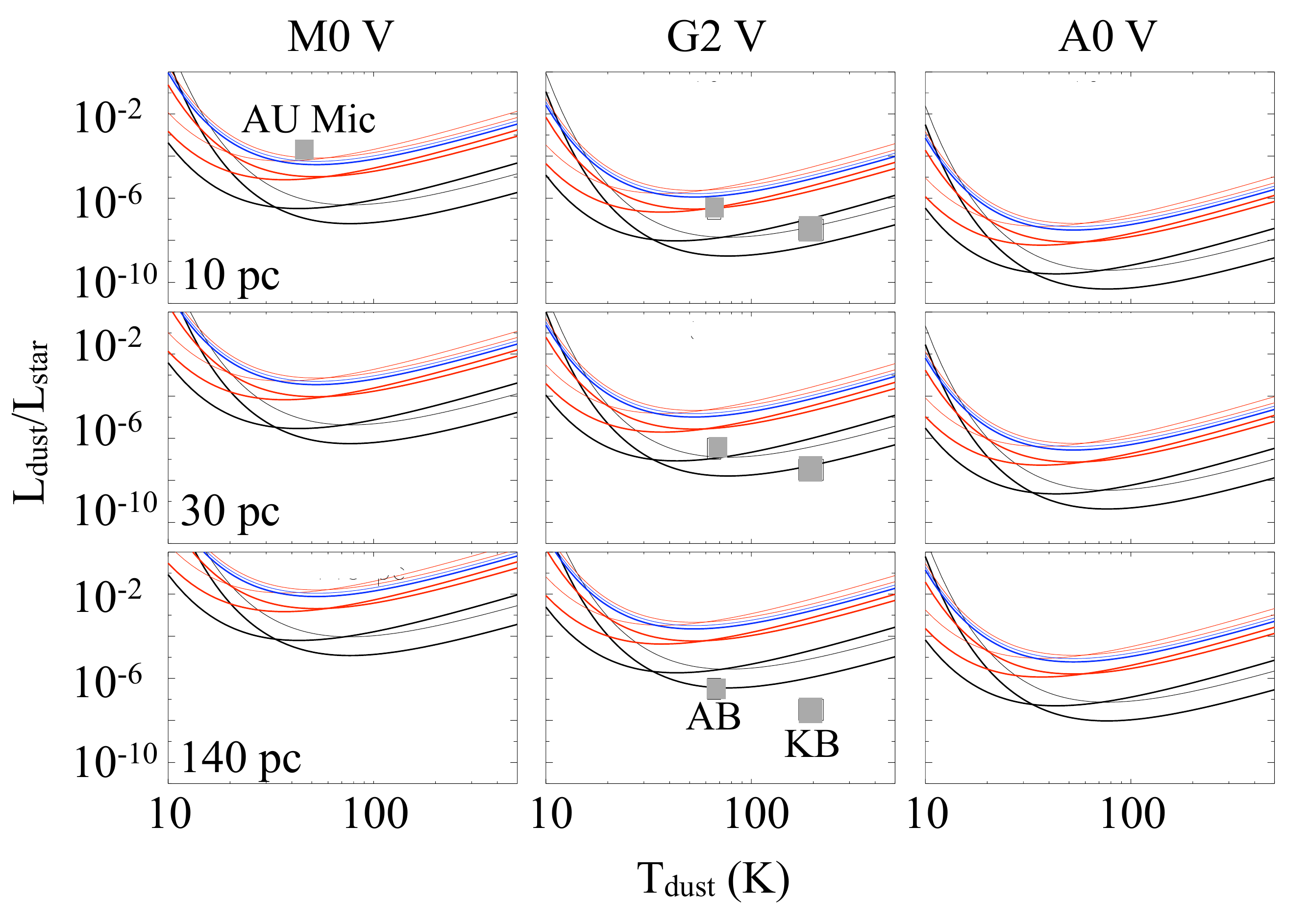}
  \caption{Minimum detectable fractional luminosities (L$_{dust}$/L$_*$) for debris disks at 10 pc ({\bf top}), 30 pc ({\bf middle}) and 140 pc ({\bf bottom}), around stars of different spectral types: M0 V ({\bf left}), solar G2 V ({\bf middle} -- with characteristic values for the Kuiper and asteroid bets shown in grey) and A0 V ({\bf right}). The different colors correspond to different instruments with the following sensitivities (3-$\sigma$ in 1 hour).  
{\it Spitzer/MIPS} at 70 $\mu$m: 1.1 mJy  but confusion limited to 6 mJy ({\bf blue}); 
{\it Herschel/PACS} at 70 $\mu$m and at 100 $\mu$m: 1.6 mJy and 1.7 mJy, respectively ({\bf red});
{\it SPICA/SAFARI} at 48 $\mu$m: 6.6 $\mu$Jy ({\bf black}); {\it SPICA/SAFARI}  at 85 $\mu$m: 7.8 $\mu$Jy  but confusion limited to 60 $\mu$Jy ({\bf black}). The sensitivity is assumed to scale as $t^{-1/2}$. The {\bf thick} lines are for  3-$\sigma$ detections in 1 hour, while the {\bf thin} lines are for 3-$\sigma$ detections in 1 minute. Note: these figures assume that the stellar photosphere can be subtracted to a high level of precision.
}
\label{sensitivity}
\end{figure}

Figure \ref{sensitivity} shows the minimum detectable fractional luminosities (L$_{dust}$/L$_*$) for debris disks at different distances as observed with 
{\it Spitzer/MIPS} at 70 $\mu$m, {\it Herschel/PACS} at 70 $\mu$m and 100 $\mu$m, and {\it SPICA/SAFARI} at 48 $\mu$m and 85 $\mu$m. {\it Herschel} will improve {\it Spitzer}'s performance because its larger aperture helps to decrease the confusion limit, but their sensitivities are comparable. Contrarily,  {\it SPICA},  being a cryogenic telescope, will have a much better sensitivity that will allow to detect  colder and fainter debris disks (the latter either due to distance or small dust content -- see the are enclosed by the dotted lines in Figure \ref{histo}). 
For solar-type stars, Figure \ref{sensitivity} shows that {\it Spitzer/MIPS} could not detect dust at the Kuiper or asteroid belt levels even for the closest disks and,  as we mentioned above, the debris disk detections at 70 $\mu$m were generally limited to $\sim$100$\times$ the luminosity of the dust in the Kuiper belt.
Contrarily, the figure shows that {\it SPICA/SAFARI} will be able to detect much fainter and distance disks, allowing to reach the level of dust in the KB out to distances of 140 pc (assuming the stellar photosphere can be subtracted to a high level of precision). This can represent a revolution of debris disks studies because of the large number of debris disks harboring stars that may be detected: as we mentioned above, models and observations indicate that debris disks might be very common and there are $\sim$10$^{5}$ F0--K2 stars within 140 pc, a number to be compared to the $\sim$ 190 debris disks around these type of stars known to date.  {\it SPICA/SAFARI} high sensitivity photometry can therefore increase the number of debris disks detections around solar-type stars by 3 orders of magnitude, allowing better statistics of the  disk frequencies and properties as a function of stellar type, age and environment. 
In addition, 140 pc is the distance to the Taurus star forming region, and this opens a new discovery space for {\it SPICA/SAFARI}: the study of disks in the transition from a protoplanetary  to a debris stage. Another leap forward will be the study of debris disks around M-type stars,  particularly interesting because these stars are the most common ($\sim$ 77\% in the local neighborhood), and there is observational evidence that at least 3\% harbor planets and that super-earths are more common than giants. Figure \ref{sensitivity} shows that debris disks orders of magnitude fainter than that around AU-Mic (at 9.9 pc) could be detected, and there are $\sim$ 160 M-type stars  (with 0.3--0.8 $M_{\odot}$) within that distance.  

Planetesimals are the building blocks of planets. {\it SAFARI} observations of debris disks will shed light on whether planetesimals are present around most stars, i.e. on how  robust is the process of planetesimal formation.  Also of critical importance is that {\it SAFARI} spectrophotometry (35--210 $\mu$m) will be able to trace the shape and the peak of the SED (see Figure \ref{sed}), allowing to constrain the dust characteristic temperature and location (rather than assuming that the SED peaks at 70 $\mu$m, implying  $T_{dust}$ $\sim$ 50 K, as it was done with most debris disks observed by {\it Spitzer} that were not detected beyond that wavelength). This will allow to address whether KB-like planetesimals are ubiquitous, important because these icy planetesimals can be reservoirs of water that could potentially be delivered to the "habitable zones" of these extra-solar planetary systems. 

\small
\begin{center}
3. DYNAMICAL EVOLUTION OF EXTRA-SOLAR PLANETESIMAL BELTS
\end{center}
\normalsize

The fractional luminosity ($L_{dust}/L_*$) of the KB and AB dust in the solar system today is  $\sim$ 10$^{-7}$--10$^{-6}$ and 10$^{-8}$--10$^{-7}$, respectively, but it is thought that the solar system was much dustier in the past because the AB and KB were much more populated\footnote{Evidence of this is that the present KB is not sufficiently populated to have formed objects with diameter $>$ 200 km by pairwise accretion in 4 Gyr, and that minimum mass solar nebula needed to account for the solids found in the solar system bodies shows a strong depletion in the AB region that is unlikely primordial.}. The distribution of sizes of the asteroids and KBOs shows evidence that the primordial planetesimal belts have suffered significant erosion, but erosion alone can not account for the inferred degree of planetesimal depletion. The cratering record of the terrestrial planets and the Moon indicate that this depletion was likely the result of the dynamical scattering that took place around the time of the Late Heavy Bombardment (LHB). This event, dated from lunar samples of impact melt rocks, happened during a very narrow interval of time, 3.8 to 4.1 billion years ago, in which most of the impact craters were formed; thereafter the impact rate decreased exponentially. The impact crater record shows that the LHB lasted 20-200 Myr, that the source of the impactors was the main AB, and that the mechanism for this event was the orbital migration of the giant planets which caused a resonance sweeping of the AB, and a large scale ejection of asteroids into planet-crossing orbits \cite{gomes2005}, \cite{strom2005}. The orbital migration of the planets also caused a depletion of the KB as Neptune migrated outward. The LHB was probably a single event in the history of the solar system that would have been accompanied by a high rate of asteroid collisions and dust production, which would have caused a large spike in the warm dust luminosity of the solar system \cite{booth2009}. 

\small
\begin{center}
{\it Contribution of SPICA}
\end{center}
\normalsize

The LHB was caused by the migration of the giant planets and the resulting gravitational perturbations on the asteroid and Kuiper belts. Because there is evidence of both ingredients, namely planet migration and planetesimal belts, exist in extra-solar planetary systems, a natural question arises: are LHB-type of events (i.e. a drastic depletion of planetesimals preceded by a brief period of increased dust production)  common in other planetary systems? {\it Spitzer} observations seemed to indicate that they are not common \cite{booth2009}, however,  as mentioned above, these observations were sensitivity limited. With its superb sensitivity, {\it SPICA/SAFARI}  will allow for deeper surveys that  will be able to increase the number of debris disks detections by several orders of magnitude, improving the study of the dust emission as a function of stellar age.  This will shed light on the dynamical evolution of the system (setting constrains on planet migration), and on the frequency and timing of heavy bombardment-type of events  well after the period of planet and planetesimal formation, both of which have important consequences for the habitability of these extra-solar planetary systems. 

\small
\begin{center}
3. COMPOSITION OF EXTRA-SOLAR PLANETESIMALS
\end{center}
\normalsize

\begin{figure}
  \includegraphics[angle = 0, height=0.33\textheight]{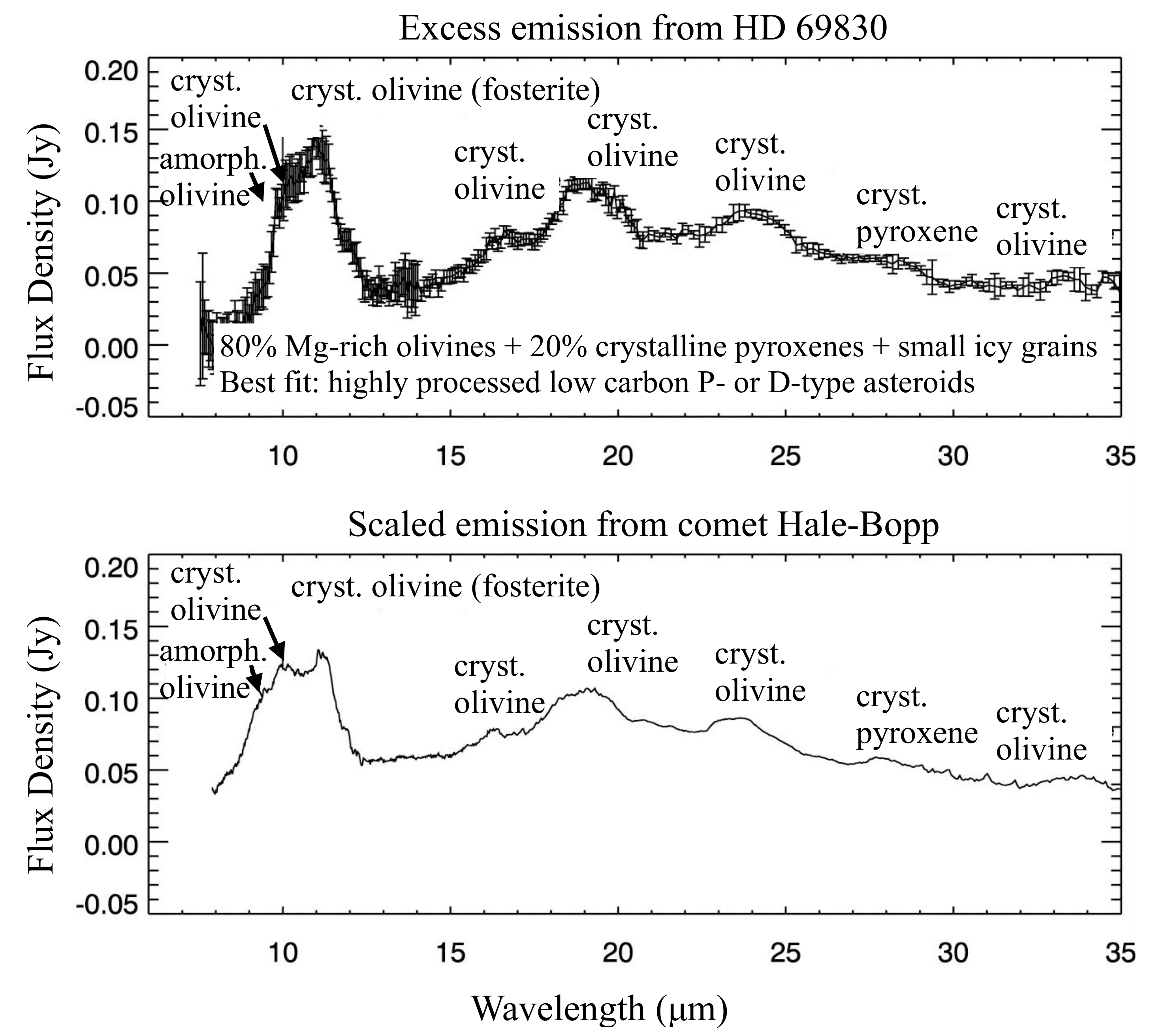}
  \caption{Spectrum of the dust excess emission from HD 69830 (top) compared to
the spectrum of comet Hale-Bopp normalized to a blackbody temperature of 400 K
(bottom). From \cite{beichman2005}.}
\label{spectra}
\end{figure}

An example of a debris disks system with an unusually high dust production rate is HD 69830,  an old K0V star (0.8 M$_{\odot}$, 0.45 L$_{\odot}$,  $>$ 2 Gyr)  known to harbor three Neptune-like planets inside 0.63 AU, that shows a strong dust emission at 24 $\mu$m but none at 70 $\mu$m \cite{beichman2005}, indicating that the dust is warm and is located close to the star.  In this and other debris disk systems like  BD+20307, HD 72905 and $\eta$-Corv, the large quantities of dust, together with the short lifetime of the small dust grains, indicate that the inferred high dust production rates could not have been sustained for the age of the system (because the dust-producing planetesimals would not survive), implying that the dust production processes are transient \cite{wyatt2007}.  These transient disks might be be linked to planet migration (like in the case of the LHB) or to large individual collisional events. In the case of HD 69830, the spectrum of the dust excess shows strong silicate emission lines remarkably similar to the emission of comet Hale-Bopp [see Figure \ref{spectra} from \cite{beichman2005}]. Further analysis revealed that the best fit to the spectrum is that of small grains of highly processed material similar to that of a disrupted P- or D-type asteroid plus small icy grains,  likely located outside the outermost planet \cite{lisse2007}. P and D-type asteroids are the most common in the outer asteroid belt, and about 25\% of the debris dust in the inner solar system comes from the break up of one of those asteriods (in an event that happened $\sim$8 Myr ago whose fragments formed the Veritas asteroidal family).
  
\small
\begin{center}
{\it Contribution of SPICA}
\end{center}
\normalsize

Debris disk spectroscopy can constrain the composition of the dust-producing planetesimals in extra-solar planetary systems. However, most debris disks observed by  {\it Spitzer/IRS} did not show any spectral features at 8--35 $\mu$m (like those in Figure \ref{spectra}). Instead, the majority of the disks' SEDs showed a smooth blackbody continuum. This might be due to the low signal-to-noise of the observations or to the grains being larger than $\sim$ 10 $\mu$m, the latter implying that the emission would be seen at $\lambda$ $>$ 35 $\mu$m, outside the  {\it Spitzer/IRS} spectral range.  {\it SPICA} will be able to help on both fronts. On the one hand,  {\it SPICA/SAFARI} spectroscopy in the 35--210 $\mu$m range will allow to study the mineralogy of 100s of disks, and for the closest targets, it will be possible to trace the variation in dust mineral content and grain size distribution as a function of disk radius that can be compared to composition of asteroids and KBOs in the solar system (also to be studied by {\it SPICA}); the presence of amorphous and crystalline water ice features in this wavelength range -- some of which have already been observed in protoplanetary disks --  opens the possibility that these spatially resolved spectroscopic observation may be able to detect the location of the snow line. On the other hand, 
{\it SPICA}’s coronograph at 3--27 $\mu$m will  increase the signal-to-noise ratio of the debris disks observations, helping in the detection of silicate and ice features; for distant targets, it will allow the imaging and spectroscopy of the inner parts of the disk. 

\small
\begin{center}
4. SPATIALLY RESOLVED DEBRIS DISKS
\end{center}
\normalsize

Most of the $\sim$ 300 debris disks known to date are spatially unresolved and little is known about their morphologies, as only limited information can be derived from the study of their SEDs.  The characteristic dust temperature inferred from the SED is in the range 50--150 K, corresponding to dust located at 10s--100 AU, however, the SED analysis is degenerate and depends on assumptions of grain size and composition.  As we explain below, the location of the dust is critical to interpret some of the debris disks observations. 

\small
\begin{center}
{\it Contribution of SPICA}
\end{center}
\normalsize

It is expected that planet and planetesimal formation processes lead to large quantities of dust  due to gravitational perturbations produced by large 1000 km-sized planetesimals that excite the orbits of a swarm of 1--10 km-size planetesimals, increasing their rate of mutual collisions \cite{kenyon2005}. This warm dust can  serve as a proxy of planetesimal formation in the terrestrial planet region. For solar type stars, the 24 $\mu$m emission traces the 4--6 AU region. Based on the incidence rate of 24 $\mu$m excesses, \cite{meyer2008} argued that if the dust-producing events are very long-lived, the stars that show dust excesses in one age bin will also show dust excesses at later times, in which case the frequency of warm dust (and terrestrial planet formation) is  $<$ 20\%.  However, if the dust-producing events are transient, shorter than the age bins, the stars showing excesses in one age bin are not the same as the stars showing excesses at other age bins, i.e. they can produce dust at different epochs, and in this case the overall frequency of warm dust is obtained from adding all the frequencies in all age bins, which results in $>$ 60\% (assuming that each star only has one epoch of high dust production). If this is the case, the frequency of planetesimal formation in the terrestrial planet region would be high \cite{meyer2008}. However, the interpretation of the data would change if  the  observed 24 $\mu$m excesses arise from the steady erosion of cold-KB-like disks \cite{carpenter2009}.  Spatially resolved observations with the {\it SPICA}'s coronograph, able to directly locate the dust, would help resolve this issue. 

In addition, {\it SPICA/SAFARI} high angular resolution will allow to increase the number of spatially resolved disks, breaking the SED degeneracy between dust location and composition.   At 48 $\mu$m, the expected angular resolution is 3.5''; a 100 AU disk would have this angular size at $\sim$ 28 pc; the number of stars within that distance is approximately 210 (A), 1100 (F0--K2) and 3600 (K2--M). Similarly,  at 85 $\mu$m, the expected angular resolution is  6''; a 100 AU disk would have this angular size at $\sim$ 17 pc and the number of stars within that distance is approximately 45 (A), 230 (F0--K2) and 750 (K2--M). The bottom line is that the number of debris disks that will be resolved with  {\it SPICA/SAFARI} will greatly increase the number of spatially resolved debris disks known to date, approximately two dozen. 

\small
\begin{center}
5. DEBRIS DISK STRUCTURE AS A PLANET DETECTION METHOD
\end{center}
\normalsize

\begin{figure}
  \includegraphics[angle = 0, height=0.22\textheight]{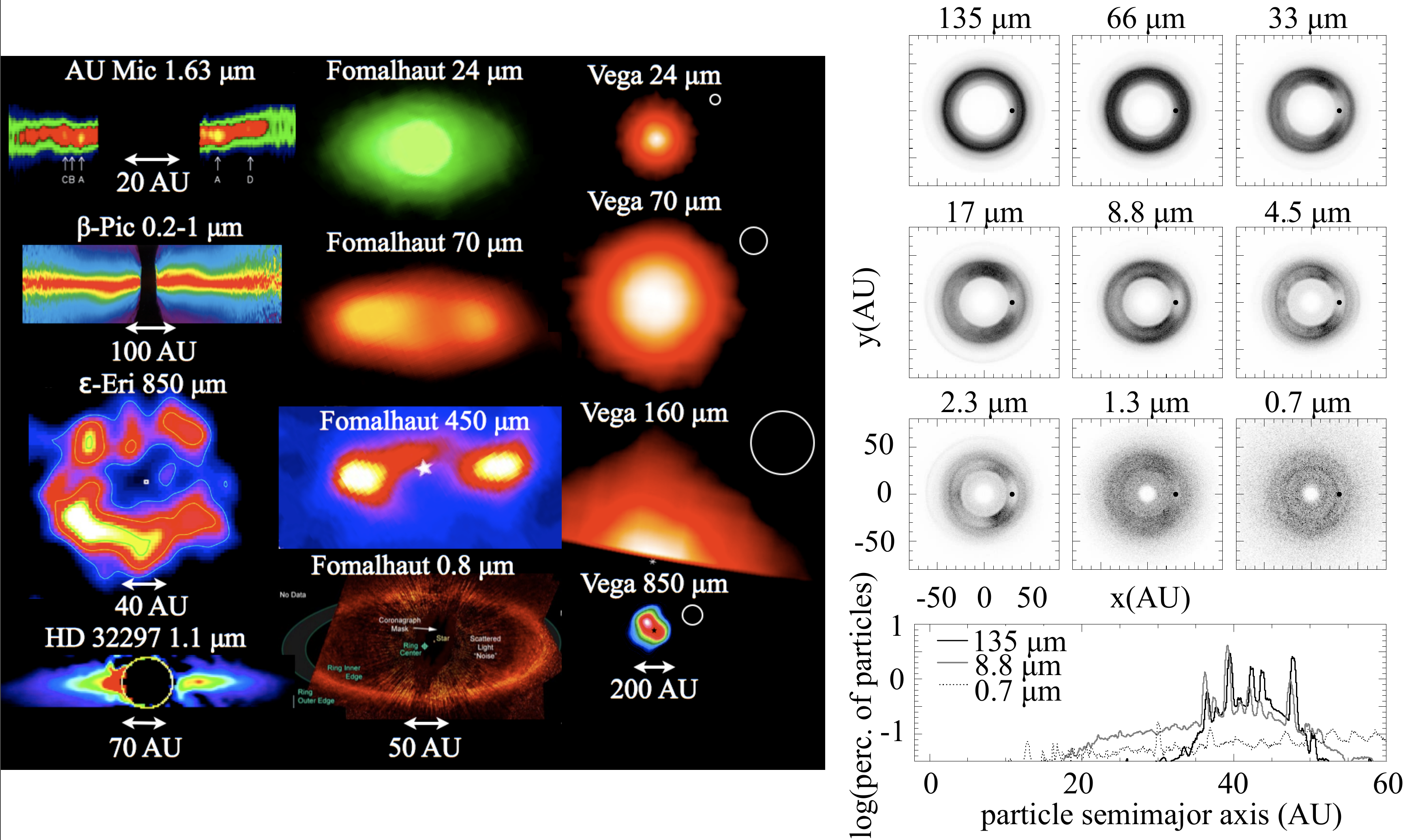}
  \caption{
  {\bf Left:} Spatially resolved debris disks showing a diversity of morphological features at different wavelengths. From top to bottom, left to right references are: \cite{liu2004}, \cite{heap2000}, \cite{greaves2005}, \cite{schneider2005}, \cite{stapelfeldt2004}, \cite{stapelfeldt2004}, \cite{holland2003}, \cite{kalas2005}, \cite{su2005}, \cite{su2005}, \cite{su2005}, \cite{holland1998}.
{\bf Right:} Expected number density distribution of the KB dust disk for nine different particle sizes (label on top of each panel). The trapping of particles in MMRs with Neptune is responsible for the ring-like structure, the asymmetric clumps along the orbit of Neptune, and the clearing of dust at Neptune's location (indicated with a black dot). The disk structure is more prominent for larger particles because the P-R drift rate is slower and the trapping is more efficient. The disk is more extended in the case of small grains because small particles are more strongly affected by radiation pressure. The histogram shows the relative occurrence of the different MMRs for different sized grains, where the large majority of the peaks correspond to MMRs with Neptune. The inner depleted region inside $\sim$ 10 AU is created by gravitational scattering by Jupiter and Saturn. Panel adapted from \cite{ama2002}. 
  }
\label{structure}
\end{figure}

Figure \ref{structure} (left) shows some spatially resolved debris disks. The disks look different at different wavenlengths because, as we saw before, they trace different dust temperatures and therefore dust locations (with the warmer dust emitting at shorter wavelengths located closer to the star -- see e.g. the case of Fomalhaut); in addition, different wavelengths trace different particles sizes, and different particle sizes have different dynamical evolutions that result in different features : large particles dominate the emission at longer wavelengths, they interact weakly with the stellar radiation field and therefore their location might resemble that of the dust-producing planetesimals, while small grains dominate at shorter wavelengths and interact more strongly with radiation, giving rise to more extended disks (clearly illustrated in the case of Vega). 

These observations show a rich diversity of structural features, including warps (AU-Mic \& $\beta$-Pic), offsets of the disk center with respect to the central star ($\epsilon$-Eri and Fomalhaut), brightness asymmetries (HD 32297 and Fomalhaut), clumpy rings (AU-Mic, $\beta$-Pic, $\epsilon$-Eri and Fomalhaut) and sharp inner edges (Fomalhaut). Some of these features could be due to gravitational perturbations of massive planets and in some cases they have been observed in the solar system debris disk.   The basic mechanisms by which the gravitational perturbations produced by a massive planet can create structure in the debris disk are the trapping  in mean motion resonances (MMRs), the effect of secular resonances and gravitational scattering. Dust particles drifting inward under P-R drag can become entrapped in exterior MMRs because at these locations the particle receives energy from the perturbing planet that can balance the energy loss due to P-R drag, halting their inward migration. In systems where the planet migrates outward this can also result in the trapping of planetesimals (like e.g. the plutinos in the solar system trapped in the 3:2 MMR with Neptune).  Because the particles spend most of their lifetime trapped in exterior MMRs, this results in the formation of resonant rings outside the planet's orbit and, depending on the geometry of the resonance, in the formation of clumps. After the particles leave the resonance and continue to drift inward, they can get ejected from the system by gravitational scattering. Gravitational scattering is very efficient and can result in formation of dust depleted regions inside the planet's orbit \cite{ama2005}.  Figure \ref{structure} (right) shows the effect of resonant trapping and gravitational scattering on the KB dust disk.  Secular perturbations, the long-term average of the perturbing forces, act on timescales $>$ 0.1 Myr, can be very long ranged, and can result in the formation of an inner dust depleted region, a warp (if the planet and the planetesimal disk are not coplanar or if there are two non-coplanar planets), and/or an  offset in the disk center and resulting brightness asymmetry (if the planet is in an eccentric orbit -- \cite{wyatt1999}).  Secular perturbations are responsible for the inner edge of the main asteroid belt and for the offset, brightness asymmetry and warp observed the inner solar system debris disk. 

\small
\begin{center}
{\it Contribution of SPICA}
\end{center}
\normalsize
 
The debris disks structure resulting from these gravitational perturbations depends on the mass and location of the planet and is sensitive to medium-to-large sized planets with a wide range of semimajor axis (out to 100s of AU). Therefore, the study of debris disk structure with {\it SPICA's}, enabled by its high spatial resolution, will explore a new parameter space in planet detection not covered by  radial velocity and transit surveys (limited to planets close to the star) and by direct imaging (limited to young planets). The study of the presence of long period planets is important to study the diversity of planetary systems, to constrain planet formation models, and to establish the stability of planets in the habitable zone. In addition, {\it SPICA's} coronograph will allow to search for planets and dust on similar spatial scales, shedding light on the planet-disk interaction. \\
\\
{\it The study of debris disks with {\it SPICA} can shed light on the diversity of planetary systems, the link between debris disks and planets and the link between extra-solar planetary systems and our own.}

\begin{acknowledgements}
 A.M.M. is a Ram\'on y Cajal Fellow, funded by the  Spanish Government. 
\end{acknowledgements}

\end{document}